\documentclass[a4paper]{article}

\usepackage{amsthm,amsbsy,amsfonts,amssymb,color,amsmath,psfrag,array,hyperref}
\usepackage{enumerate}
\usepackage{minibox}
\usepackage{setspace}
\usepackage{commath}
\usepackage{multirow}
\usepackage{mathtools}
\usepackage{amsthm}
\usepackage{booktabs}
\newcommand{\tabitem}{~~\llap{\textbullet}~~}

\title{Learning from Data to Optimize Control in Precision Farming}

\author{Alexander Kocian $^{1}$* and Luca Incrocci $^{2}$}

\date{%
$^{1}$ \quad Department of Computer Science, University of Pisa, 56127 Italy; \\
$^{2}$ \quad Department of Agriculture, Food and Environment, University of Pisa, 56124  Italy; Luca.incrocci@unipi.it}

\begin{document}

\maketitle

\begin{abstract}
Precision farming is one way of many to meet a 70 percent increase in global demand for agricultural products on current agricultural land by 2050 at reduced need of fertilizers and efficient use of water resources. The catalyst for the emergence of precision farming has been satellite positioning and navigation followed by Internet-of-Things, generating vast information that can be used to optimize farming processes in real-time. Statistical tools from data mining, predictive modeling, and machine learning analyze pattern in historical data, to make predictions about future events as well as intelligent actions. This special issue presents the latest development in statistical inference, machine learning and optimum control for precision farming.
\end{abstract}





\section{Introduction}

The world’s population is expected to be nearly 10 billion by 2050, corresponding to a 55 percent increase in global demand for agricultural production based on current trend. In 2011, according to FAO, agriculture made use of 2710 km$^3$ (70 percent) of all water withdrawn from aquifers, stream and lakes, but this number masks large geographical discrepancies. Middle East, Northern Africa and Central Asia, has already withdrawn most of the exploitable water with 80–90 percent of that going to agriculture. Hence, rivers and aquifers are depleted beyond sustainable levels \cite{FAO-11}. Shifting the focus to arable land, 1.6 billion hectares are arable worldwide. The total world land area suitable for cropping is 4.4 billion corresponding to around 40 percent of world's land. However, in several regions, soil quality constraints affect more than half the cultivated land base, notably in sub-Saharan Africa, Southern America, Southeast Asia and Northern Europe \cite{FAO-11}. When forests are converted into farming land, the largest stores of carbon locked in those trees will be released to the atmosphere, contributing to global warming on top of today's level.

Clearly, crop production on current land needs to be increased through adopting new technologies. To increase profits, reduces waste and maintains environmental quality at the same time, farmers are supplied with decision support systems that propose the right dose/action at the right place and at the right time \cite{Gallardo-20,Thompson-20}. The core piece of such decision support system is a agricultural model related to either crop growth, epidemiology, or market development that optimizes a control function based on probabilistic assessment of causal relationships \cite{Shi-19}. Satellite telemetry tracking data along with existing geo-referenced digital map as well as Internet-of-Things based sensor data act as input to the model. Automated data processing systems, often located in the cloud, train the model. The trend goes from manually trained to self-calibrating models that adapt to changes in the environment over time. Smartphone applications have become a key interface in precision agriculture between the farmer and the cloud. These applications not only visualize the control parameters and suggests possible actions but also return the farmers' reaction (irrigation, sowing, fertilization etc.) back to the cloud. Fully automatized actions that go beyond human level performance while minimizing resources are still subject to research.

\section{Statistical Inference and Machine Learning}

The key to effective experimentation in precision farming is blocking, replication, and randomization \cite{Young-96}. To analyze and interpret the experimental results as well as to predict upcoming data, tools from statistics are deployed. Probabilistic models approximate the complex dynamics of the underlying process using statistical assumptions on the generation of sample data. Statistics draws population inferences from data samples. Neither training and nor test sets are necessary to infer the parameters. The supervised machine learns from training data to build a statistical model that can be used to make repeatable predictions. The unsupervised machine, in contrast, learns the model on its own without external training data. With the development of Internet-of-Things, machine learning applications for precision farming have been rapidly developing over the last years \cite{Liakos-18}.

\subsection{Low-Order Statistics}

Random variables have a discrete or continuous probability distribution. Low-order statistics denote the first and second moments of a sample from the distribution. The former and the second correspond to the mean and the statistical auto and cross power of the random variables. Low order statistics, however, require a very large number of samples to estimate with a reasonable level of confidence. When the random variables are Normal distributed, this now ranked data is often used for \emph{ANalysis Of VAriance} (ANOVA), comparing the ratio of within group variance and between group variance, to assess systematic factors (bias) and random factors (covariance). The former has statistical influence on the data set while the latter does not. For example, there is an average weight variation within one kind of pumpkin but there might be another average weight variation among different pumpkin varieties. The \emph{Pearson correlation coefficient} defines as ratio of covariance to the product of individual variances measures linear correlation between two random variables. For example, the Pearson correlation between evapotranspiration and precipitation is positive over the southern/deforested but negative over the northern/forested Amazonia \cite{Vergopolan-16}.

\subsection{Regression}

Multiple regression models characterize the relationship between a dependent target variable and multiple weighted independent feature variables. The weights, also known as regression coefficients, are an average functional relationship between target and features which might be linear or non-linear. For example, an exponential regression is adequate to model the relation between tree height and leaf-area index of Prunus \cite{Pardossi-09}. The least square fitting technique yields the model parameters. A probit regression, in contrast, considers binary target variables with Gaussian distributed model noise and possibly multiple weighted independent variables. The maximum likelihood technique is often used to obtain the model parameters. Voting with binary outcome is a typical application of probit regression. For example, Sevier and Lee used this method in \cite{Sevier-04} to predict the probability of Florida citrus producers adopting precision agriculture technologies. Note that regression analysis is sensitive to multicollinearity, arising whenever two or more independent variables used in a regression are strongly correlated with each other. In this case, the weights become very sensitive to small changes in the model.

An Artificial Neural Networks (ANNs) consists of many simple connected nodes dubbed neurons, each deploying a real-valued non-linear activation function. Input neurons are activated by data from external sensors. Other neurons are activated by weighted edges from previously active neurons. Feed-forward neural networks, forming a directed acyclic graph, process the sensed data without memory. In contrast, the recurrent neural network (RNN) allows connections among neurons in the same or previous layers. They have internal memory and their graph is directed with cycles. When fed with environmental and historical dynamic information, this type of neural network is well-suited to time series forecasting \cite{Balducci-18}. In the convolutional neural network (CNN), forward and backward propagations perform convolutional operations. Usually, the edge weights are point estimates based on stochastic gradient training. Bayesian Neural Networks model the uncertainty of the estimated edge weights by interpreting them as maximum likelihood or maximum a posteriori estimates. A comprehensive state-of-the-art overview of ANN is available in \cite{Schmidhuber-14}. Notable examples in precision farming are the feedforward neural network by Adisa \emph{et al.} in \cite{Adisa-19} for maize production prediction. In this work the feature space is spanned by the environmental parameters,  potential evapotranspiration, soil moisture and land cultivated. Barbosa \emph{et al.} deployed in \cite{Barbosa-20} a CNN that predicts the spatial yield map of corn fields in Illinois, Nebraska and Kansas, USA. Here, satellite images as well as environmental data span the feature space. In a third notable application, multi-layer (deep) CNN have been applied in \cite{Bapat-20} to detect plant leaf diseass based on 54000 (large number) training images. Finally, we want to point out the example in \cite{Chen-15} where RNN has been used for spatio-temporal prediction of leaf area index in rubber plantation. The feature space in the experiment was spanned by the individual CCD images. The underlying theory of many neural network architectures is still in research phase.

Bayesian time-series forecasting is another promising field of research in precision farming. Within this framework, all sources of uncertainty are expressed by stochastic processes. The Bayes Theorem turns the \emph{a priori} probability and the distribution of the observed data, also known as likelihood, into the posteriori distribution of the parameters for predictive inference. A partially observed state-space model such as the Hidden Markov Model (for discrete states) or the Kalman filter (for continuous states) are ideally suited to describes the dynamics of the process. A typical example in agriculture research is price prediction of crops. In \cite{Cenas-17}, a Kalman filter has been deployed to predict the price time-series of rice. When the model parameters are unknown, the observation sequence and the state sequence can be used to estimate them. The linear dynamic Bayesian network developed in \cite{Kocian-20/2} does this by relating indicative parameters of crop development to environmental control parameters. The expectation-maximization algorithm is used to track the states in the expectation step and to learn the parameters of the Bayesian network in the maximization. At iterative convergence, the algorithm provides a time-series predictor many time instants ahead. When the dynamics is non-linear on top of that, sequential Monte Carlo techniques often lead to accurate parameter predictions by sampling from the posterior distribution on the expenses of computational complexity. In the special case of sigmoid-type growth dynamics, a linear dynamic model leads to the exact predictor for the reciprocal time-series of the parameter \cite{Kocian-20/4}.

\subsection{Classification}

Classification is a supervised learning problem as above regression is. Considering models for solving classification problems, the classical Fisher linear discriminant analysis is a standard multivariate technique both for dimension reduction and supervised classification. The data vectors are transformed into a low dimensional subspace the maximize separation of class centroids. In many applications, however, the linear boundaries do not adequately separate the classes. Roth and Steinhage present in \cite{Roth-99} a nonlinear generalization of discriminant analysis that uses the kernel trick to replace dot products with an equivalent kernel function. 

Sparse Kernel Machines evaluate the kernel function only at a subset of the training data points to predict a new data point, making the computation time feasible \cite{Bishop-09}. Specifically, the support vector machine (SVM) by Boser \emph{et al.} in \cite{Boser-92} discards all data points but the support vectors, once the model is trained. The determination of the model parameters is an convex optimization problem so that any local solution is also a global solution in contrast to many other algorithms. The SVM has become popular for solving problems in classification, regression and novelty detection. For example, Jheng \emph{et al.} predicted in \cite{Jheng-18} the rice yield in Taiwan by a SVM using training data from 1995-2015. The relevance vector machine (RVM) \cite{Tipping-00} is a Bayesian sparse kernel technique that provides posterior probability outputs in contrast to the SVM. At the same time, RVM based prediction models utilise dramatically fewer basis function than a comparable SVM. To name an example from remote sensing, the RVM with plate spline kernel is able to spatially estimate chlorophyll from an unmanned aerial system at low computational cost \cite{Elarab-15}. Finally, we want to point out the Informative Vector Machine (IVM), constructing sparse Gaussian process classifiers by greedy forward selection with criteria based on information theoretic principles. The IVM performs similar to the SVM by only a fraction of training data. Roscher \emph{et al.} uses in \cite{Roscher-12} an incremental version of the IVM to classify hyperspectral image data for various agricultural crops in Italy, Europe, and Indiana, USA.

\subsection{Clustering}

Clustering is an unsupervised process of partitioning a set of data (or objects) in a set of meaningful sub-classes, called clusters. Clustering techniques can be categorized into i) partitioning algorithms constructing various partitions and then evaluate the result by some criterion (k-means, k-medoids, CLARANS,...); ii) hierarchical algorithms creating a hierarchical decomposition of the set of objects by some criterion (AGNES, BIRCH, CURE, DIANA,...); iii) density-based methods that are guided by connectivity and density functions (DBSCAN, OPTICS,...); iv) grid-based methods that are based on a multi-level granularity structure (STING, WaveCluster, CLIQUE, ...); and v) model-based clustering methods that find the best fit to a hypothetical model (Autoclass, Rock, EM-algorithm,...). Massive computing power makes it possible, for example, to mine large amount of existing crop, soil and climatic data. Clustering the result based on districts with maximum wheat yield gives the optimal range of best temperature, worst temperature and rain fall \cite{Majumdar-17}. To scale clustering algorithms with the number of dimensions and the number of data items, attention has been drawn to distributed approach \cite{Hore-04}. Nevertheless, the scaling problem is still a challenge for most of above clustering algorithms such as big data applications.

\section{Closing the Loop}

\begin{table}[!h]
\begin{tabular}{|p{4cm}|p{5cm}|p{5cm}|}
\hline
Method & Strengths & Weaknesses \\ \hline \hline
MANOVA & \parbox{5cm}{  \tabitem Powerful test for finding truly significant factors. \\ \tabitem Robust to Type I errors. } & \parbox{5cm}{  \tabitem Relation between independent grouping variable and dependent variables sometime ambiguous. \\ \tabitem Computationally complex. } \\ \hline
Multiple Regression & \parbox{5cm}{  \tabitem Theory well understood. \\ \tabitem Good results are obtained with relatively small data sets. \\ \tabitem Ability to determine impact of independent variable on dependent variable. } & \parbox{5cm}{  \tabitem Missing data erroneously changes regression coefficients.\\ \tabitem Correlation does not necessarily correspond to a causation.\\ \tabitem Sensitive to outliers.  }\\ \hline
Deep Neural Networks & \parbox{5cm}{  \tabitem Perform well on audio, image, text data. \\ \tabitem Architecture can be adapted to a number of problems. } & \parbox{5cm}{  \tabitem Computationally intensive to train. \\ \tabitem Tuning hyper-parameters needs expert knowledge.  }\\ \hline
Dynamic Bayesian Network & \parbox{5cm}{  \tabitem Accurate prediction of temporal behavior. \\ \tabitem Flexible adapts to environmental changes. \\ \tabitem Underlying theory is well understood. } & \parbox{5cm}{  \tabitem Cannot handle real biological systems with feedback loops (cycles). \\ \tabitem Initial guess of parameters is crucial for convergence.  }\\ \hline
Support Vector Machine  & \parbox{5cm}{  \tabitem Memory efficient. \\ \tabitem Flexible (non-linear) threshold using Kernels. \\ \tabitem Convex optimization problem with unique solution. } & \parbox{5cm}{  \tabitem Does not scale with data dimension. \\ \tabitem Sensitive to tuning the regularization parameters (overfitting).\\ \tabitem  Finding a proper kernel is often cumbersome. }  \\ \hline

k-means clustering & fast, simple. & Model order must be known in advance. \\ \hline
DBSCAN clustering & \parbox{5cm}{  \tabitem Model-order free. \\ \tabitem Scalable. \\ item Estimate is unbiased. } & \parbox{5cm}{  \tabitem Sensitive to choice of hyperparameters. \\ \tabitem Good results only for uniform densities. } \\ \hline
Reinforcement Q-Learning & \parbox{5cm}{  \tabitem Computes most successful rewards even when the environment is large. \\ \tabitem Model-free. \\ \tabitem Convergence to the optimum policy is guaranteed. }  & \parbox{5cm}{  \tabitem Computationally complex. \\ \tabitem Assumes that all of the states and all of the actions are presentable as matrix.  } \\
\hline
\end{tabular}
\caption{Comparison of common statistical models and machine learning algorithms.}
\label{tab:comparison}
\end{table}

So far, machines have mostly be used to learn from the observations with the goal to predict future outcome given current conditions. Clearly with increasing number of observations, the machine becomes smarter over time but it does not have control over the environmental conditions. Currently, these are controlled by the agronomist's experience. A more efficient approach is to let agents make optimal actions subject to minimizing resources. The result is a close-loop precision farming system where the model learns from data in the forward loop and controls actuators in the backward loop, as outlined in \cite{Burchi-18}. Reinforcement learning, making smarter decisions over time, has enjoyed a great success in several domains such as computer game, medical diagnosis and energy management. Bu and Wang build in \cite{Bu-19} a smart agriculture IoT system based on deep reinforcement learning that decides the amount of water needed to be irrigated by analyzing the collected agricultural environment data. Though there had been great progress, the technology cannot yet achieve the human-level performance in adaptation to dynamic environments and solving complex tasks. Ergo, there is still a lot of space for research towards optimum precision farming. Table~\ref{tab:comparison} lists strengths and weaknesses of common statistical models and machine learning algorithms.

\section{Conclusions}

Precision farming for current arable land is a promising approach to meet the vast global demand for agricultural products on current land. Internet-of-Things provides vast real-time information on crop related parameters, soil and weather, that feeds machine learning algorithms for better crop productivity while protecting the environment. The ultimate goal is to maximize yield by minimizing water consumption, usage of fertilizers, and amount of arable land in an automatic fashion. Although there has been an evolution of research in this area, more knowledge is needed to close the gap between current practice and optimum precision farming.


%






\bibliographystyle{IEEEtran}
\bibliography{../../../database/agriculture,../../../database/IT,../../../database/mypubs,../../../database/economy}

\begin{thebibliography}{10}
\providecommand{\url}[1]{#1}
\csname url@samestyle\endcsname
\providecommand{\newblock}{\relax}
\providecommand{\bibinfo}[2]{#2}
\providecommand{\BIBentrySTDinterwordspacing}{\spaceskip=0pt\relax}
\providecommand{\BIBentryALTinterwordstretchfactor}{4}
\providecommand{\BIBentryALTinterwordspacing}{\spaceskip=\fontdimen2\font plus
\BIBentryALTinterwordstretchfactor\fontdimen3\font minus
  \fontdimen4\font\relax}
\providecommand{\BIBforeignlanguage}[2]{{%
\expandafter\ifx\csname l@#1\endcsname\relax
\typeout{** WARNING: IEEEtran.bst: No hyphenation pattern has been}%
\typeout{** loaded for the language `#1'. Using the pattern for}%
\typeout{** the default language instead.}%
\else
\language=\csname l@#1\endcsname
\fi
#2}}
\providecommand{\BIBdecl}{\relax}
\BIBdecl

\bibitem{FAO-11}
FAO, \emph{The state of the world's land and water resources for food and
  agriculture}, FAO, Ed.\hskip 1em plus 0.5em minus 0.4em\relax Rome, Italy:
  How should agriculture produce enough food for the world?, 2011.

\bibitem{Gallardo-20}
M.~Gallardo, A.~Elia, and R.~B. Thompson, ``Decision support systems and models
  for aiding irrigation and nutrient management of vegetable crops,''
  \emph{Agricultural Water Management}, no. 106209, 2020.

\bibitem{Thompson-20}
R.~B. Thompson, L.~Incrocci, J.~van Ruijven, and D.~Massa, ``Reducing
  contamination of water bodies from {E}uropean vegetable production systems,''
  \emph{Agricultural Water Management}, no. 106258, 2020.

\bibitem{Shi-19}
X.~Shi, X.~An, Q.~Zhao, H.~Liu, L.~Xia, X.~Sun, and Y.~Guo, ``State-of-the-art
  {I}nternet of {T}hings in protected agriculture,'' \emph{MDPI Sensors},
  vol.~19, no. 1833, 2019.

\bibitem{Young-96}
J.~C. Young, ``Blocking, replication, and randomization - the key to effective
  experimentation,'' \emph{Quality Engineering}, vol.~9, no.~2, pp. 269--277,
  1996.

\bibitem{Liakos-18}
K.~Liakos, P.~Busato, D.~Moshou, S.~Pearson, and D.~Bochtis, ``Machine learning
  in agriculture: A review,'' \emph{Sensors}, vol.~18, no.~8, p. 2674, 2018.

\bibitem{Vergopolan-16}
N.~Vergopolan and J.~B. Fisher, ``The impact of deforestation on the
  hydrological cycle in {A}mazonia as observed from remote sensing,''
  \emph{Tayler \& Francis Int. Journal of Remore Sensing}, vol.~37, no.~22, pp.
  5412--5430, 2016.

\bibitem{Pardossi-09}
A.~Pardossi, L.~Incrocci, G.~Incrocci, F.~Tognoni, and P.~Marzialetti, ``What
  limits and how to improve water use efficiency in outdoor container
  cultivation of ornamental nursery stocks,'' \emph{ISHS Acta horticulturae},
  pp. 73--80, 10 2009.

\bibitem{Sevier-04}
B.~J. Sevier and W.~S. Lee, ``Precision farming adoption by florida citrus
  producers: {P}robit model analysis,'' in \emph{Proc. ASABE Annual Meeting},
  no. Paper No. 041080, Ottawa, Canada, 8 2004.

\bibitem{Balducci-18}
F.~Balducci, D.~Impedovo, and G.~Pirlo, ``Machine learning applications on
  agricultural datasets for smart farm enhancement,'' \emph{Machines}, vol.~6,
  no.~3, p.~38, 2018.

\bibitem{Schmidhuber-14}
J.~Schmidhuber, ``Deep learning in neural networks: An overview,'' University
  of Lugano \& SUPSI, Switzerland, Tech. Rep. IDSIA-03-14, 10 2014.

\bibitem{Adisa-19}
O.~Adisa, J.~Botai, A.~Adeola, A.~Hassen, C.~Botai, D.~Darkey, and
  E.~Tesfamariam, ``Application of artificial neural network for predicting
  maize production in {S}outh {A}frica,'' \emph{MDPI Sustainability}, vol.~11,
  no.~4, 2 2019.

\bibitem{Barbosa-20}
A.~Barbosa, R.~Trevisan, N.~Hovakimyan, and N.~F. Martin, ``Modeling yield
  response to crop management using convolutional neural networks,''
  \emph{Computer and Electronics in Agriculture}, vol. 170, 3 2020.

\bibitem{Bapat-20}
A.~Bapat, S.~Sabut, and K.~Vizhi, ``Plant leaf disease detection using deep
  learning,'' \emph{SERSC Int. J Advanced Science and Technology}, vol.~29,
  no.~6, pp. 3599--3605, 2020.

\bibitem{Chen-15}
B.~Chen, Z.~Wu, J.~Wang, J.~Dong, L.~Guan, J.~Chen, K.~Yang, and G.~Xie,
  ``Spatio-temporal prediction of leaf area index of rubber plantation using
  {HJ-1A/1B CCD} images and recurrent neural network,'' \emph{ISPRS Journal of
  Photogrammetry and Remote Sensing}, vol. 102, pp. 148--160, 2015.

\bibitem{Cenas-17}
P.~V. Cenas, ``Forecast of agricultural crop price using time series and kalman
  filter method,'' \emph{Asia Pacific Journal of Multidisciplinary Research},
  vol.~5, no.~4, pp. 15--21, 2017.

\bibitem{Kocian-20/2}
A.~Kocian, D.~Massa, S.~Cannazzaro, L.~Incrocci, S.~D. Lonardo, P.~Milazzo, and
  S.~Chessa, ``Dynamic {B}ayesian network for crop growth prediction in
  greenhouses,'' \emph{{ELSEVIER} Computer and Electronics in Agriculture},
  Feb. 2020.

\bibitem{Kocian-20/4}
A.~Kocian, G.~Carmassi, C.~Fatjon, L.~Incrocci, P.~Milazzo, and S.~Chessa,
  ``Bayesian sigmoid-type time series forecasting with missing data for
  greenhouse crops,'' \emph{MDPI Sensors}, vol.~20, no.~11, Jun. 2020.

\bibitem{Roth-99}
V.~Roth and V.~Steinhage, ``Nonlinear discriminant analysis using kernel
  functions,'' \emph{Adv. Neural Information Processing Systems}, pp. 568--574,
  12 1999.

\bibitem{Bishop-09}
C.~M. Bishop, \emph{Pattern Recognition and Machine Learning}.\hskip 1em plus
  0.5em minus 0.4em\relax Springer Science + Business Media LLC, 2009.

\bibitem{Boser-92}
B.~E. Boser, I.~M. Guyon, and V.~N. Vapnik, ``A training algorithm for optimal
  margin classifiers,'' in \emph{Proc.~5th ACM Annual Workshop on Computation
  Learning Theory (COLT'92)}, Pittsburgh, PA, USA, 7 1992, pp. 144--152.

\bibitem{Jheng-18}
T.-Z. Jheng, T.-H. Li, and C.-P. Lee, ``Using hybrid {S}upport {V}ector
  {R}egression to predict agricultural output,'' in \emph{Proc. 27th Wireless
  and Optical Comm. Conf. (WOCC 2018)}, Hualien, Taiwan, 2018.

\bibitem{Tipping-00}
M.~E. Tipping, ``The relvance vector machine,'' \emph{Advances in Neural
  Information Processing Systems}, vol.~12, pp. 652--658, 2000.

\bibitem{Elarab-15}
M.~Elarab, A.~Ticlavilca, A.~Torres-Rua, I.~Maslova, and M.~McKee, ``Estimating
  chlorophyll with thermal and broadband multispectral high resolution imagery
  from an unmanned aerial system using relevance vector machines for precision
  agriculture,'' \emph{ELSEVIER Int. J. Appl. Earth Observation and
  Geoinformation}, vol.~43, pp. 32--42, 2015.

\bibitem{Roscher-12}
R.~{Roscher}, B.~{Waske}, and W.~{Forstner}, ``Incremental import vector
  machines for classifying hyperspectral data,'' \emph{IEEE Transactions on
  Geoscience and Remote Sensing}, vol.~50, no.~9, pp. 3463--3473, 2012.

\bibitem{Majumdar-17}
J.~Majumdar, S.~Naraseeyappa, and S.~Ankalaki, ``Analysis of agriculture data
  using data mining techniques: application of big data,'' \emph{Springer
  Journal of Big Data}, vol.~4, no.~20, 2017.

\bibitem{Hore-04}
P.~Hore and L.~O. Hall, ``Scalable clustering: A distributed approach,'' in
  \emph{Proc.~IEEE Int. Conf. on Fuzzy Systems}, Budapest, Hungary, 2004, pp.
  143--148.

\bibitem{Burchi-18}
G.~Burchi, S.~Chessa, F.~Gambineri, A.~Kocian, D.~Massa, P.~Milano, P.~Milazzo,
  L.~Rimediotti, and A.~Ruggeri, ``Information technology controlled
  greenhouse: A system architecture,'' in \emph{Proc.~IoT Vertical and Topical
  Summit for Agriculture}.\hskip 1em plus 0.5em minus 0.4em\relax Tuscany,
  Italy: IEEE, May 2018.

\bibitem{Bu-19}
F.~Bu and X.~Wang, ``A smart agriculture {IoT} system based on deep
  reinforcement learning,'' \emph{ELSEVIER Future Generation Computer Systems},
  vol.~99, pp. 500--507, 10 2019.

\end{thebibliography}





\end{document}